\documentclass{webofc}
\usepackage[varg]{txfonts}   
\woctitle{Physics at the magnetospheric boundary}
\begin{document}
\title{Latest results of pulse phase resolved spectroscopy of cyclotron lines in accretion powered pulsars}
%
%

\author{Chandreyee Maitra\inst{1,2}\fnsep\thanks{\email{cmaitra@rri.res.in
    }} \and
        Biswajit Paul\inst{1}\fnsep\thanks{\email{bpaul@rri.res.in}} 
}

\institute{Raman Research Institute,Sadashivnagar, Bangalore-560080, India
\and
           Joint Astronomy Programme, Indian Institute of Science, Bangalore-560094.
          }

\abstract{%
  We have performed pulse phase resolved spectroscopy of the Cyclotron Resonance Scattering Features
 (CRSF) of some bright accretion powered X-ray pulsars like 1A 1118-61, Vela X-1, A0535+26, XTE J1946+274,
 4U 1907+09, 4U 1626-67 and GX 301-2 using \emph{Suzaku} observations with long exposures. We have performed
 the study using different spectral models for the continuum and have 
obtained similar patterns of variations of the CRSF in all the cases, thus demonstrating the robustness of 
our results. Pulse phase dependence of the CRSF in XTE J1946+274 has been obtained for the first time, and phase resolved
variations of the CRSF in 4U 1907+09 has been compared at factor of $\sim$ 2 difference in luminosity. We have also
studied the pulse profiles of these objects near the CRSF energy, and have noticed an increased pulse fraction and/or
a change in
the pulse shape near the CRSF energy for some sources. The implications of the results are discussed.
}
\maketitle
\section{Introduction}
\label{intro}
Cyclotron Resonance Scattering Features
(CRSF) which have been discovered in about 20 sources till now provide us with
a wealth of information regarding accretion powered pulsars. Assuming the CRSF is formed on the surface of the 
neutron star, the centroid energy directly converts to the magnetic field strength of the neutron star by the 
12 $B_{12}$ rule : $\ensuremath{E_{c}}=11.6 \, keV\times\frac{1}{1+z}\times\frac{B}{10^{12}G}$, \ensuremath{E_{c}} being the
centroid energy, z the gravitational redshift and B the magnetic field strength of the neutron star. Pulse phase 
dependence of the CRSF parameters further provides important clues regarding the emission geometry
at different viewing angles and can also map the magnetic field geometry of the neutron star. 
As the scattering cross sections are significantly altered and increased by a large factor near the cyclotron
resonance scattering feature \citep{araya1999,araya2000,schonherr2007}, the corresponding
beam patterns and hence the pulse profiles are also expected to change near the
corresponding resonance energies of the accretion powered pulsars . Such probes have
been attempted before by  \cite{tsygankov2006,lutovinov2009,ferrigno2011}, where a change in the beaming pattern has been detected near the CRSF energy.
\emph{Suzaku} with its broad energy range and high sensitivity is ideal for pursuing
the above studies and we have made very detailed investigations of the pulse phase dependence of the CRSF,
and their beaming pattern near the CRSF energy for some sources which are bright in nature and were 
observed with \emph{Suzaku} with long exposures.\\
\section{Observations \& Analysis}
\label{sec-1}
We reduced and processed the \emph{Suzaku} observations of several bright accretion powered pulsars with a known CRSF, and
which had observation duration $\ge 50$ ks.
For the transient sources like 1A 1118-61, A 0535+26 and XTE J1946+274 data were obtained near an outburst
of the respective source. Pile-up check was done and data from the inner pixels were rejected
when required . The energy spectra
were modelled phenomenologically with different continuum models used to fit the spectra of
accretion powered pulsars like the powerlaw with high energy cutoff (highecut), or powerlaw with the Fermi Dirac 
cutoff (fdcut), the cutoff powerlaw (cutoffpl) model, the negative-positive
exponential powerlaw component (NPEX),
or the more physical comptonization model 'CompTT'. Partial covering absorber was added for sources with low energy dips
in their pulse profiles, and iron emission lines were added whenever required. We tried to fit the energy spectra
with all the continuum models mentioned above, available as a standard or local 
package in \emph{XSPEC} and carried out further analysis with only the model(s) which
gave best fits for the respective sources. For the CRSF we used a Lorentzian profile "cyclabs" in
\emph{XSPEC}.
\section{Methods and checks used in the analysis}
(1) Careful modeling of the broadband continuum spectrum and obtaining consistent
and physically reliable parameters of the continuum for the different models. (2) The CRSF parameters should give consistent results for both the Gaussian and
the Lorentzian profile. (3)The stretch of the observation chosen for the analysis should be free from significant
luminosity or spectral variability.

%
%
\section{Results}
\subsection{Phase resolved spectroscopy of CRSF}
\begin{figure}[ht!]
\centering
\includegraphics[height=4 cm, width=3cm,angle=-90]{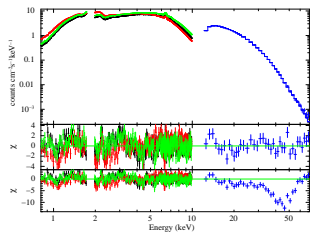}
\caption{\scriptsize Phase averaged spectrum of 1A 1118-61 with the best fit continuum model NPEX (first panel) and residuals
before (second panel)
and after (third panel) inclusion of the CRSF. Note that our previous result published on this source (MNRAS, 2012, 420, 2307)
required a very broad CRSF
with the 'cutoffpl' model. We have modified our results with the 'NPEX' and 'FDCUT' models which gives a much narrower
and better constrained CRSF, and have also modified the results of pulse phase resolved spectroscopy similarly.}
\label{fig1}
\end{figure}

\begin{figure}[ht!]
\begin{center}$
\begin{array}{l c c r}
\includegraphics[height=6 cm, width=3cm]{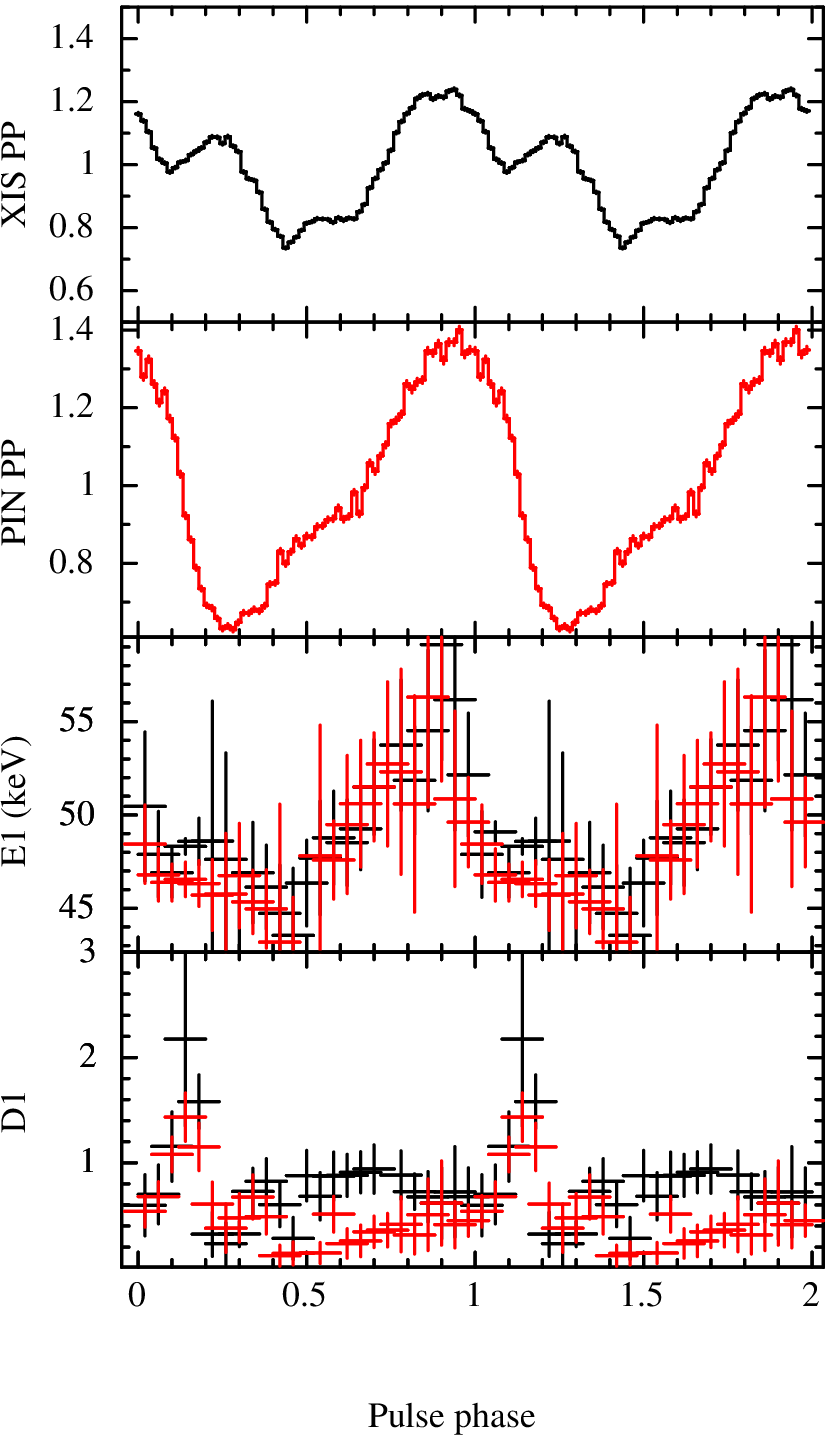} &
\includegraphics[height=6 cm, width=3cm]{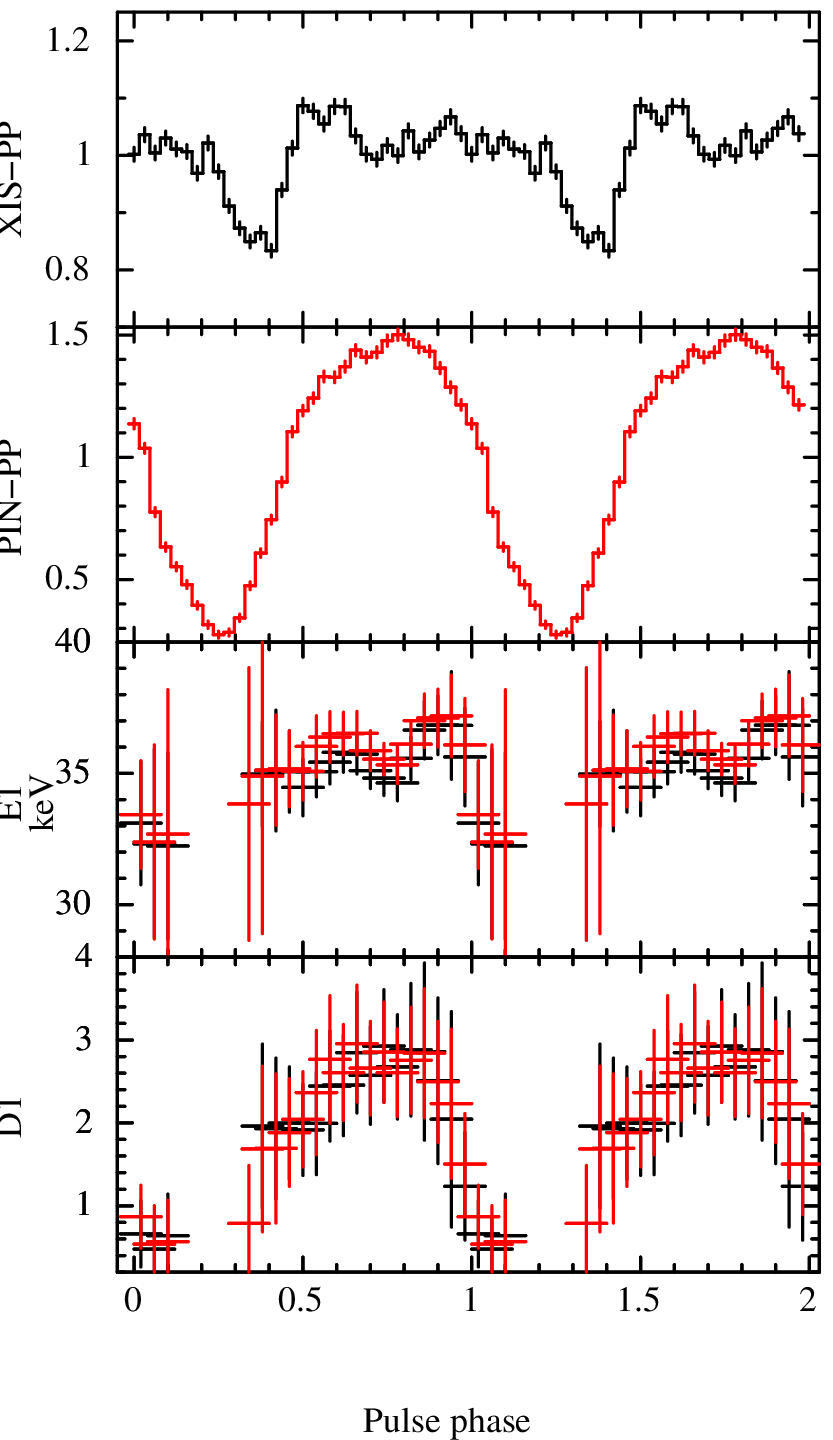} &
\includegraphics[height=6 cm, width=3cm]{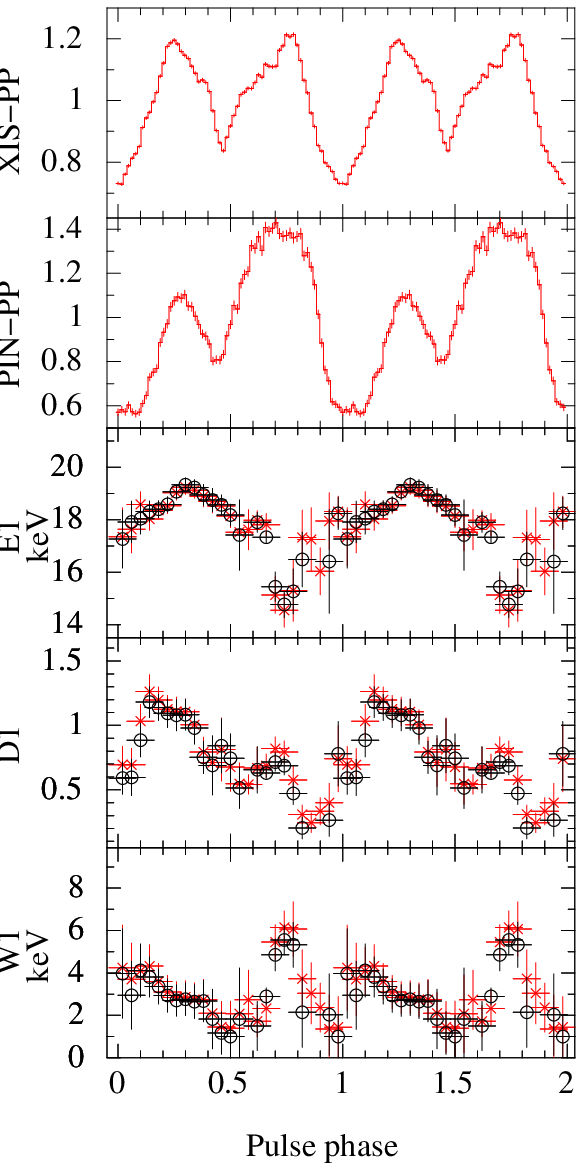} &
\includegraphics[height=6 cm, width=3cm]{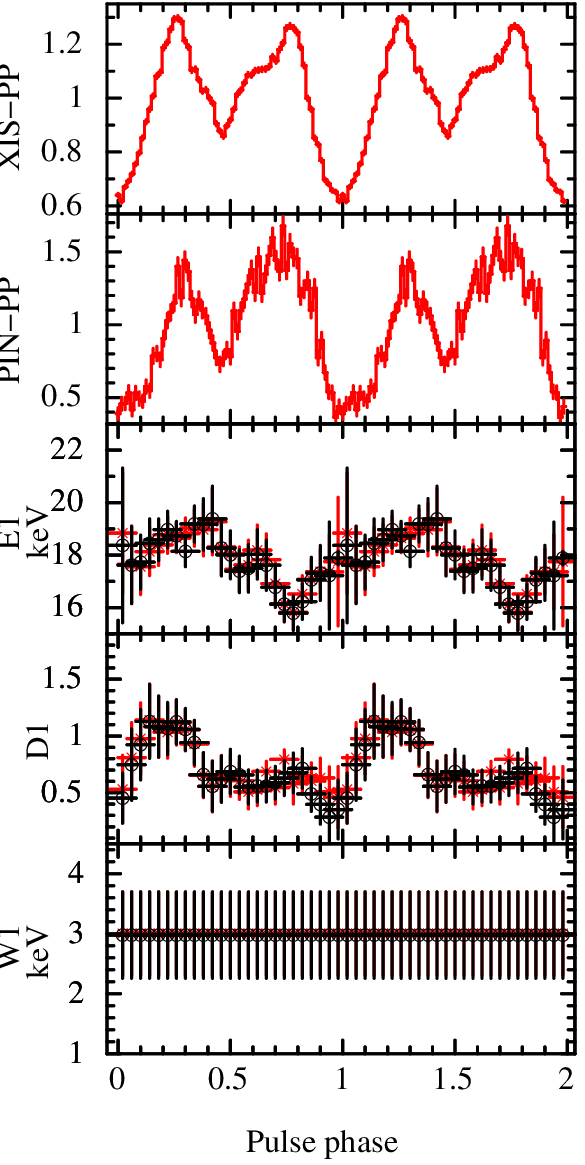} \\
\end{array}$
\end{center}
\caption{\scriptsize{The leftmost panels denote the pulse phase resolved spectroscopy of CRSFs in 1A 1118-61
\& 4U 1626-67 respectively. The next two denote the same for 4U 1907 at
factor of $\sim$ 2 difference in luminosity (third from left for the brighter observation).
In all the figures, the top two panels indicate normalized intensity of the XIS and PIN profiles
respectively, and the 'red' and 'black' points
denote the results obtained from different continuum models.}}
\label{fig2}
\end{figure}

Phase resolved spectra were generated with the phases centered around 25 independent
bins but at thrice their widths. This resulted in 25 overlapping bins out of which only 8
were independent. All the individual spectra for the respective sources were fitted with
the best fit continuum and line models obtained from fitting the phase averaged spectra. For phase resolved
spectroscopy, due to limited
statistics we froze the width of the CRSF for most sources. For clarity, Figure
\ref{fig1} shows the phase averaged spectrum of a source (1A 1118-61) with the best fit continuum model before
and after the inclusion of the CRSF. 
Table \ref{table1} summarizes the results obtained for the different
sources. A part of these results are already published in \citep{maitra2013a,maitra2013b}.
Figures \ref{fig2} shows some new results from the pulse phase resolved study of the CRSFs. 
\subsection{Probing the pulse profiles near the CRSF energy}
\begin{figure}[ht!]
\begin{center}$
\begin{array}{l c r}
\includegraphics[scale=0.22]{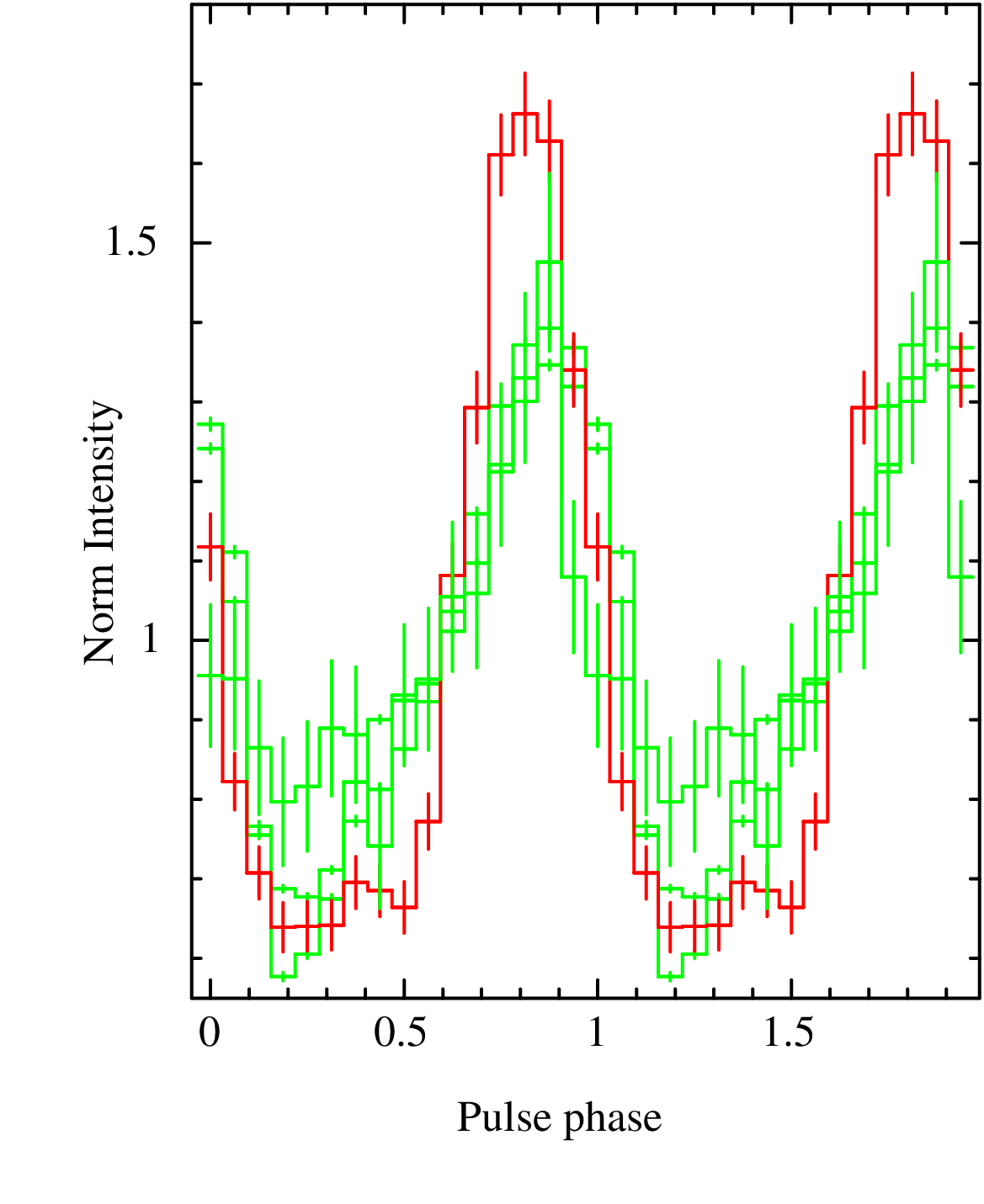} &
\includegraphics[scale=0.2]{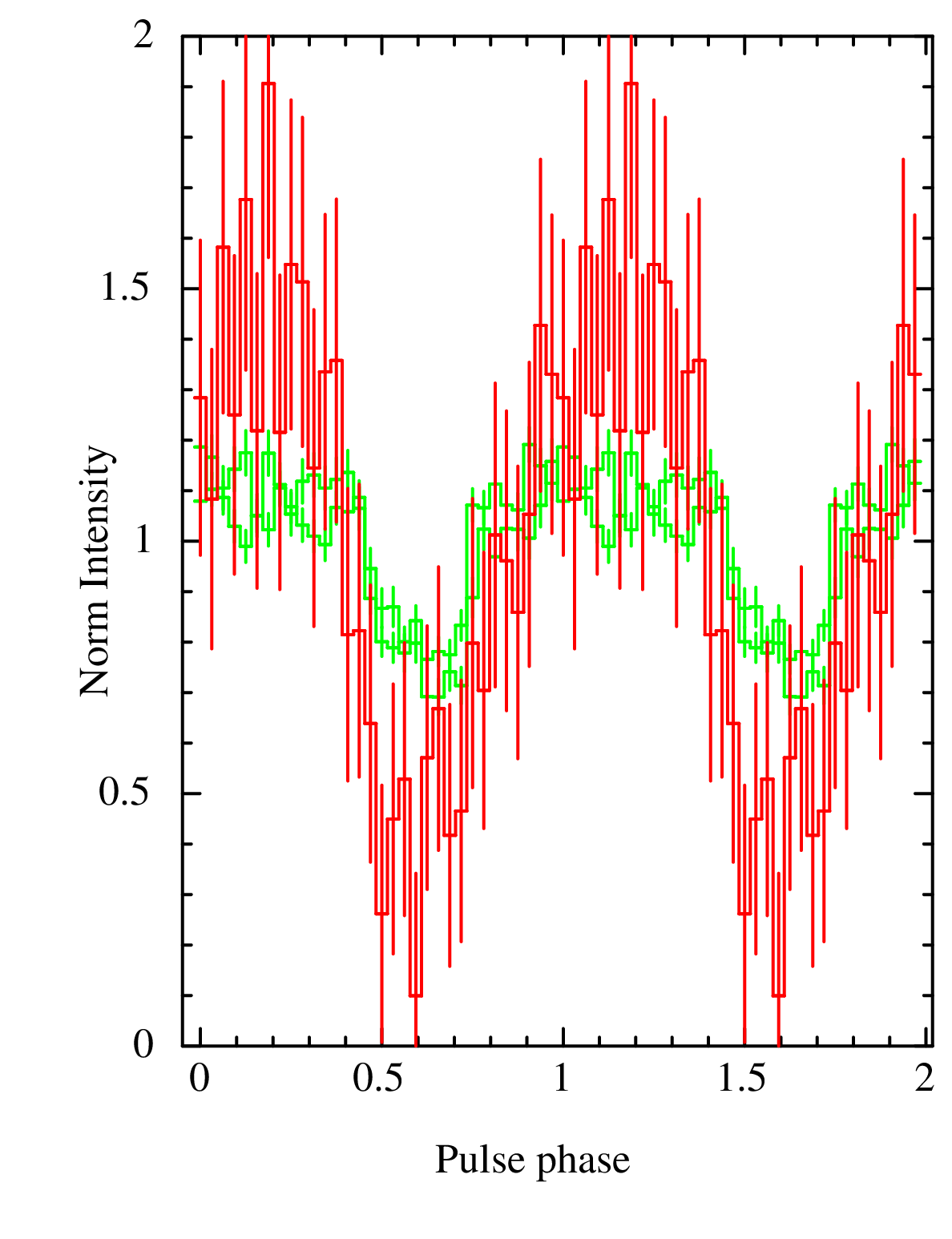} &
\includegraphics[scale=0.2]{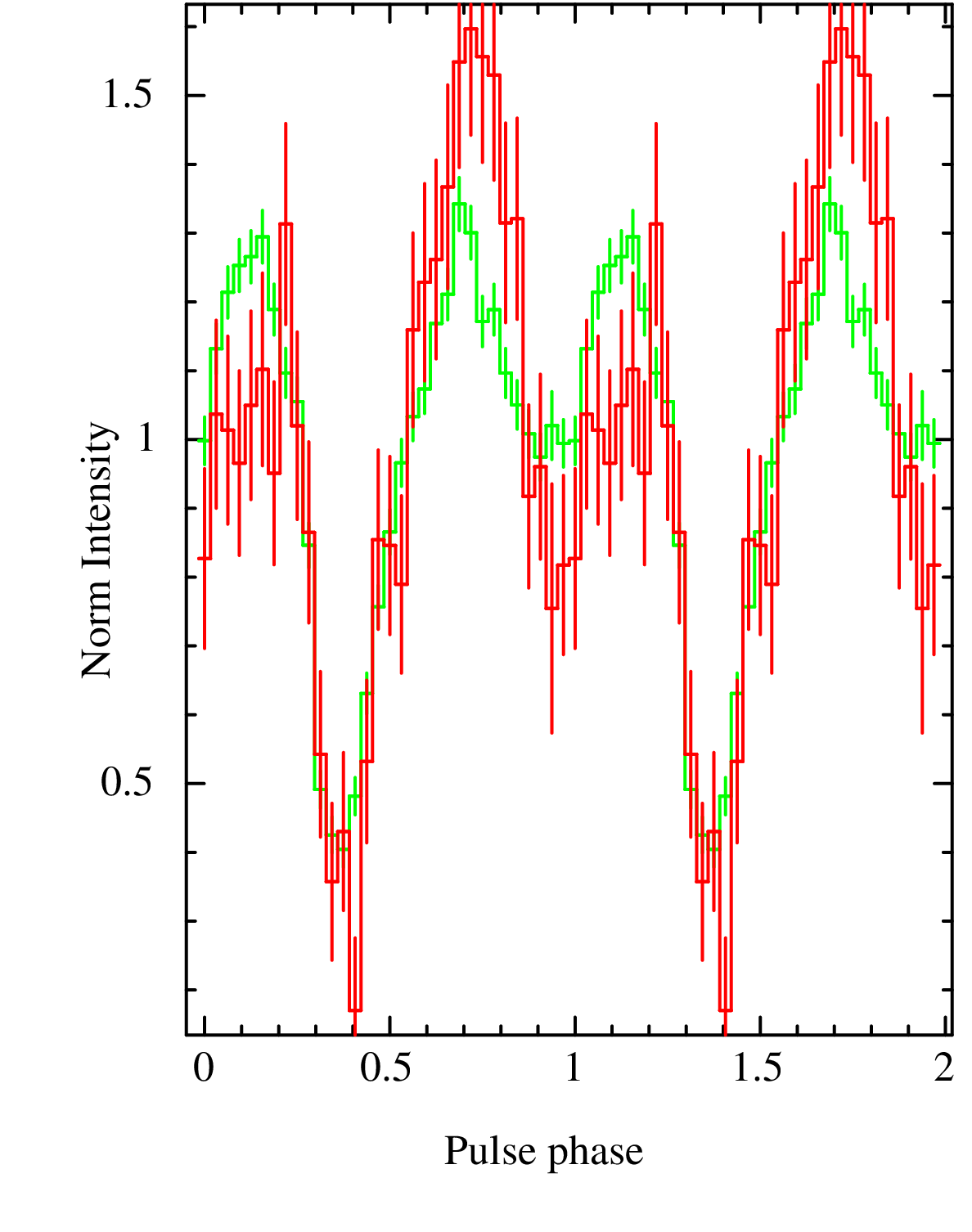} \\
\end{array}$
\end{center}
\caption{\scriptsize{Energy dependent pulse profiles of 1A 1118-61, A 0535+26 and XTE J1946+274 (left to right).
The pulse profiles near the CRSF are marked in 'red'. Profiles at other energies are marked in 'green'. Pulse profiles
denote normalized intensity.}}
\label{fig3}
\end{figure}

We have studied the energy dependence of the pulse profiles of the above sources near the CRSF energies.
1A 1118-61, A 0535+26 and XTE J1946+274 shows enhanced beaming and change in the shape of the pulse near
the CRSF energy as shown in Figure \ref{fig3}. Profiles in "Red" are at the CRSF energies and profiles
in "Green" are at energies above and below the CRSF band.
\begin{table}[h!] 
	\caption{Table on CRSF parameter variation in sources}
	\label{table1}
       \scriptsize
	\centering
        \begin{tabular}{l l l l l }
        \hline
	Source name & Best Fit Continuum models & $E_{CRSF}$ & $\%$ variation of CRSF ($E_{CRSF}$) & comments \\
	 & & keV & & \\
	\hline
	4U 1907+09 & NPEX \& CompTT & 18 & 19 & CRSF parameters show similar variation for factor of 2 \\
	& & & & difference in $L_{x}$\\
	Vela X-1 & Highecut \& CompTT & 24, 50 & 27 & Detected variation of the ratio of line energies with phase.\\
	4U 1626-67 & NPEX \& CompTT & 35 & 12 & --\\
	GX 301-2 & FDCUT \& Newhcut & 36 & 13 & Sharp gradients in the CRSF parameters detected. \\
	XTE J1946+274 & Highecut \& NPEX & 38 &36 & Obtained for the first time in this source.\\
	A 0535+26 & NPEX \& CompTT & 43 & 14 & -- \\
	1A 1118-61 & NPEX \& FDCUT & 47 & 30 & --\\
	\hline 
	
	\hline
	\end{tabular}
	\end{table}
\section{Summary}
This work can be summarized as follows: (1) The strength
of our results lie in the fact that we have obtained similar pattern of variation of the
CRSF parameters for all sources with more than one continuum model. 
(2) The variations in
the CRSF parameters with phase are typically 10–30 \% or more. Such high percentage
of variation cannot be explained by the change in viewing angle alone, and if
explained by the changing projections at different viewing angles of different parts of the
accretion column, requires a large gradient in the physical parameters across the emission
region. (3) We have found evidence of the 
pulse phase dependence of the ratio of the two line energies in Vela X-1 which indicates that the CRSF
fundamental and first harmonic forming regions are different and change with the changing viewing angles.
(4) In case of 4U 1907+09 CRSF parameters show the same pattern of
variation for the two observations having factors of two difference in luminosity. This has
important implications on the change in the beaming pattern of radiation with luminosity
and hints that the beaming pattern remains constant for this luminosity range. Similar
results obtained at different luminosity levels would be crucial to establish the change in
beaming pattern of the sources with luminosity as proposed in \cite{becker2012}. (5) A change
in the beaming pattern is detected near the corresponding CRSF energies through the pulse profiles of some sources. This
needs to be studied in more detail.

\vspace{0.5cm}
\textbf{Acknowledgments}
\scriptsize{
The authors thank the organizers, specially Dr. Enrico Bozzo for giving the opportunity
to write this proceeding. CM acknowledges the valuable comments and suggestions from Prof Dipankar Bhattacharya,
specially regarding the pulse profiles near the CRSF energies.}
\end{document}